\documentstyle[12pt,epsfig]{article}
\newcounter{multieqs}

\def\be{\begin{equation}}
\def\ee{\end{equation}}
\def\beq{\begin{eqnarray}}
\def\eeq{\end{eqnarray}}
\def\bea{\begin{eqnarray}}
\def\eea{\end{eqnarray}}
\def\la{\langle}
\def\ra{\rangle}
\def\D{\Delta}

\def\si{\sigma}
\def\te{\theta}
\def\p{\partial}

\def\pa{\partial}

\def\del{\Delta}
\def\ddel{{}^\bullet\! \Delta}
\def\deld{\Delta^{\hskip -.5mm \bullet}}
\def\dddel{{}^{\bullet \bullet} \! \Delta}
\def\ddeld{{}^{\bullet}\! \Delta^{\hskip -.5mm \bullet}}
\def\deldd{\Delta^{\hskip -.5mm \bullet \bullet}}
\def\idt{\int \!\! d\tau\ }

\def\idtds{\int \!\! \int \!\! d\tau d\sigma\ }

\begin{document}
\onecolumn
\hphantom{xxxxxxxxxxxxxxxxxxxxxxxxxxxx}
\hspace{4.5in} ITP-SB-98-61
\vskip -5mm
\hphantom{xxxxxxxxxxxxxxxxxxxxxxxxxxxx}
\hspace{4.5in} hep-th/9810119

\vglue 3cm

\begin{center}
{\Large {\bf On Mode Regularization of the Configuration Space 
Path Integral in Curved Space}}
\vspace{1cm}

Fiorenzo Bastianelli $^a$\footnote{E-mail: bastiane@unimo.it} and 
Olindo Corradini $^b$\footnote{E-mail: corradin@grad.physics.sunysb.edu}   
\\[.4cm]
{\em $^a$ Dipartimento  di Fisica, Universit\`a di Modena \\
via Campi 213-A, I-41100 Modena \\ and \\
 INFN, Sezione di Bologna\\ 
via Irnerio 46, I-40126 Bologna, Italy} \\[.4cm]
{\em $^b$ Institute for Theoretical Physics \\
State University of New York at Stony Brook \\
Stony Brook, New York, 11794-3840, USA}\\
\end{center}
\baselineskip=14pt
\vskip .7cm

\centerline{\large{\bf Abstract}}
\vspace{.25in}

The path integral representation of the transition amplitude for a 
particle moving in curved space has presented unexpected challenges since 
the introduction of path integrals by Feynman fifty years ago.
In this paper we discuss and review mode regularization of the 
configuration space path integral, and present a three loop computation
of the transition amplitude to test with success the consistency of such 
a regularization.
Key features of the method are the use of the ``Lee-Yang'' ghost fields,
which guarantee a consistent treatment of the non-trivial path integral 
measure at higher loops, and an effective potential specific to 
mode regularization which arises at the two loop order.
We also perform the computation of the transition amplitude 
using the regularization of the path integral by time discretization,
which also makes use of ``Lee-Yang'' ghost fields
and needs its own specific effective potential.
This computation is shown to reproduce the same final result as 
the one performed in mode regularization.

\newpage
\setcounter{footnote}{0}
\baselineskip=14pt
\section{Introduction}
\hspace{.5in}

The Schr{\"o}dinger equation for a 
particle moving in a 
curved space with metric $g_{\mu\nu}(x)$ has many applications ranging from
non-relativistic 
diffusion problems (described by a Wick rotated version of the 
Schr{\"o}dinger equation)
to the relativistic description of particles moving in a curved 
space-time. 
However it cannot be solved exactly for an arbitrary background metric 
$g_{\mu\nu}(x)$,
and one has to resort to some kind of perturbation theory.
A very useful perturbative solution can be  obtained by employing
a well-known ansatz introduced by De Witt \cite{one}, also known as the 
heat kernel ansatz. This ansatz makes use of a power series expansion in 
the time of propagation of the particle. The coefficients of the power
series are then determined iteratively by requiring that the 
Schr{\"o}dinger equation be satisfied perturbatively.

Equivalently, the solution of the Schr{\"o}dinger equation can be represented 
by a path integral, as shown by Feynman fifty years ago \cite{two}.
One can formally write down the path integral for the particle moving in 
curved space and check that the standard loop expansion
reproduces the structure of the heat kernel ansatz of De Witt. 
However the proper definition of the path integral in curved space 
is not straightforward. In fact it has 
presented many challenges due to complications arising from:
$i)$ the non-trivial path integral measure \cite{three},
 $ii)$ the proper discretization of the action necessary to 
regulate the path integral.
A quite extensive literature has been produced over the years addressing  
especially the latter point \cite{four}.

In this paper we short cut most of the literature and 
discuss a method of defining the path integral by employing mode
regularization as it is by now standard in many calculations done in 
quantum field theory. 
The methods extends the one employed by Feynman and Hibbs in discussing 
mode regularization of the path integral in flat space \cite{five}.
It has been introduced and successively refined  in \cite{six},
\cite{seven} and \cite{eight}
where quantum mechanics was used to compute one loop trace anomalies
of certain quantum field theories.
The key feature is to employ ghost fields to treat the non-trivial 
path integral measure as part of the action, in the spirit of Lee and Yang
\cite{three}.
These ghost fields have been named ``Lee-Yang'' ghosts and
allow to take care of the non-trivial path integral 
measure at higher loops in a consistent manner.
The path integral is then defined by expanding all fields, including 
the ghosts, in a sine expansion about the classical trajectories and 
integrating over the corresponding Fourier coefficients.
The necessary regularization is obtained by integrating
all Fourier coefficients up to a fixed mode $M$, which is eventually taken to 
infinity. A drawback of mode regularization is that it doesn't 
respect general coordinate invariance
in target space: a particular non-covariant 
counterterm has to be used in order to restore that symmetry \cite{eight}.
General arguments based on power counting (quantum mechanics can
be thought as a super-renormalizable quantum field theory) plus the fact that 
the correct trace anomalies are obtained by the use of this path integral
suggest that the mode regularization described above is consistent to 
any loop order without any additional input.

As usual when dealing with formal constructions, it is a good practice 
to check with explicit calculations the proposed scheme.
It is the purpose of this paper to present a full three loop computation 
of the transition amplitude. The result is found to be correct since it 
solves the correct Schr{\"o}dinger equation at the required loop order.
This gives a powerful check on the method of mode regularization for 
quantum mechanical path integrals on curved space.
In addition, we present our computation in such a way that it can be easily
extended and compared to the time discretization method developed in 
refs. \cite{nine},
which is also based on the use of the Lee-Yang ghosts.
This method requires its own specific counterterm (also called 
effective potential) to restore general coordinate invariance.
As expected  both schemes give the same answer.

The paper is structured as follows. In section 2 we review the method
of mode regularization and discuss the effective potential specific to 
this regularization.
In section 3 we present a three loop computation of the transition 
amplitude.  Here we make use of general coordinate invariance to select
Riemann normal coordinates to simplify an otherwise gigantic computation.
We check that the result satisfies the Schr{\"o}dinger equation
at the correct loop order.
In section 4 we extend our computation to the time discretization scheme.
This is found to compare successfully with the results previously
obtained in section 3.
Finally, in section 5 we present our conclusions and perspectives. 
In appendix A we present a technical section with a list of loop integrals
employed in the text. In particular, we
discuss how to compute them in mode regularization
as well as in time discretization regularization.

\section{Mode regularization}
\hspace{.5in}

The Schr{\"o}dinger equation for a particle of mass $m$
moving in a $D$-dimensional curved space with metric 
$g_{\mu\nu}(x)$ and coupled to a scalar potential $V(x)$ is given by
\be
i \hbar {\pa \over \pa t} \Psi = H \Psi
\ee
where
\be
H= - {\hbar^2 \over 2 m} \nabla^2  + V(x)
\ee
with $\nabla^2$ the covariant laplacian acting on scalars.
It can be obtained by canonical quantization of the model described by 
the classical action
\be
S_{cl}[x] = \int dt\  
\biggl [{m\over 2}  g_{\mu\nu}(x) \dot x^\mu \dot x^\nu - V(x)
\biggr ]
\ee
when ordering ambiguities are fixed by requiring general coordinate 
invariance in target space and requiring in addition that no scalar 
curvature term be generated by the orderings 
in the quantum potential\footnote{
One could be more general by coupling the particle also 
to a vector potential $A_\mu(x)$. It is simple to do so, since 
mode regularization will 
respect the corresponding gauge invariance \cite{seven}.
For simplicity we set $A_\mu(x) =0$ in this paper.}.
For convenience we will Wick rotate the time 
variable $t \rightarrow -it$
and set $m=\hbar=1$ to obtain the following heat equation

\be
- {\pa \over \pa t} \Psi = \biggl [- {1\over 2 } \nabla^2  + V(x)
\biggr ] \Psi
\label{schr} 
\ee
and corresponding euclidean action
\be
S[x] = \int dt\  \biggl [
{1\over 2}  g_{\mu\nu}(x) \dot x^\mu \dot x^\nu + V(x)
\biggr ].
\label{euac}
\ee
As mentioned in the introduction the heat equation 
can be solved by the heat kernel ansatz of De Witt \cite{one}:
\be
\Psi (x,y,t) = {1 \over {(2 \pi t)^{D \over 2}}} \
{\rm e}^{ - { \sigma(x,y)\over  t}}
\sum_{n=0}^\infty a_n(x,y) t^n
\label{ansatz}
\ee
which depends parametrically on the point $y^\mu$ that specifies the 
boundary condition
$ \Psi (x,y,0)=  {\delta^D(x-y) \over \sqrt{g(x)}}$.
Here $\sigma(x,y)$ is the so-called Synge world function
and corresponds to half the squared geodesic distance.
The coefficients $a_n(x,y)$ are sometimes called Seeley-De Witt 
coefficients\footnote{It is also customary to redefine the $a_n(x,y)$
by extracting a common factor $\Delta^{1\over 2}(x,y)$, where 
$\Delta(x,y)$ is a scalar version of 
the so-called Van Vleck-Morette determinant.}
and are determined by plugging the ansatz (\ref{ansatz})
into (\ref{schr}) and matching powers of $t$.

Now we want to describe in detail
how to get the solution of eq. (\ref{schr})
by the use of a path integral which employs the classical 
action in (\ref{euac}).
Following refs. \cite{six,seven,eight}
we write the transition amplitude 
for the particle to propagate from the initial point
$x^\mu_i$ at time $t_i$ to the final point
$x^\mu_f$ at time $t_f$ as follows
\be
\langle x^\mu_f,t_f | x^\mu_i,t_i \rangle  \equiv
\langle x^\mu_f|e^{-\beta H}|x^\mu_i \rangle =
\int_{_{x(-1) = x_i}}^{^{x(0)=x_f}}  \!\!\!\!\!  \!\!\!\!\!  \!\! 
\tilde {\cal D} x\ \exp \biggl [ - { 1\over {\beta }} S \biggr ]
\label{pi}
\ee
where
\bea
&&\hskip -1cm
S \equiv S[x,a,b,c] =
\int_{-1}^{0} \! \! \! d\tau \ \biggl [
{1\over 2} g_{\mu\nu}(x) (\dot x^\mu \dot x^\nu + a^\mu a^\nu +
b^\mu c^\nu) + \beta^2 \biggl ( V(x) + V_{MR}(x) \biggr ) \biggr ]
\label{quac}  \\
&& \hskip -1cm
 V_{MR} = 
{1\over 8} R  -{1\over 24} g^{\mu\nu} g^{\alpha \beta} g_{\gamma\delta}
\Gamma_{\mu\alpha}{}^\gamma \Gamma_{\nu\beta}{}^\delta 
\label{ct} \\
&& \hskip -1cm
\tilde {\cal D} x
=  {\cal D} x  {\cal D} a  {\cal D} c  {\cal D} c.
\label{meas}
\eea
For commodity we have shifted and rescaled the time parameter 
in the action,
$t= t_f+ \beta \tau$
with $\beta = t_f - t_i$, so that $-1 \leq \tau \leq 0$.
Note that the total time of propagation 
$\beta$ plays the role of the Planck constant 
$\hbar$ (which we have already 
set to one) and counts the number of loops.  
In the loop expansion generated by $\beta$ the potentials 
$V$ and $V_{MR}$ start contributing only at two loops\footnote{
Reintroducing $\hbar$ one can see that the classical potential
$V$ must be of order $\hbar^0$ while the counterterm $V_{MR}$
is a truly two loop effect of order $\hbar^2$.}.
The full action $S$ includes terms proportional to the
Lee-Yang ghosts, namely the commuting ghosts $a^\mu$ and the 
anticommuting ghosts  $b^\mu$ and $ c^\mu$.
Their effect is to reproduce a formally covariant measure:
integrating them out produces $ \tilde {\cal D} x = 
\prod (\det g_{\mu\nu}(x(\tau)))^{1/2}  d^D x(\tau) $.
As we will discuss, mode regularization destroys this formal covariance.
Nevertheless reparametrization covariance is recovered 
thanks to the effects of the potential $V_{MR}$
directly included in the action (\ref{quac}).
With precisely this counterterm the mode regulated path integral
in (\ref{pi}) solves the equation in (\ref{schr})
in both sets of variables $(x^\mu_f,t_f)$ and $(x^\mu_i,t_i)$
and with the boundary condition 
$\langle x^\mu_f,t | x^\mu_i,t \rangle  = {\delta^D(x^\mu_f-x^\mu_i) 
\over
\sqrt{g(x)}}$.

For an arbitrary metric $g_{\mu\nu}(x)$ one is
able to calculate the path integral only 
in a perturbative expansion in $\beta$
and in  the coordinate displacements $\xi^{\mu}$ about the final point
$x^\mu_f$: $\xi^\mu \equiv x^\mu_i - x^\mu_f$. 
The actual computation starts by parametrizing
\be
x^\mu(\tau) = x_{bg}^\mu(\tau) + q^\mu(\tau)
\label{para}
\ee
where $x_{bg}^\mu(\tau) $ is a background trajectory and
$q^\mu(\tau)$ the quantum fluctuations.
The background trajectory is taken to satisfy the free equations of 
motion and is a function linear in $\tau$
connecting $x^\mu_i$ to $x^\mu_f$ in the chosen coordinate system, thus
enforcing the proper boundary conditions
\be
 x_{bg}^\mu(\tau) = x_f^\mu  - \xi^\mu \tau.
\ee
Note that 
by free equations of motion we mean the ones arising from
(\ref{quac}) by neglecting the potentials $V+V_{MR}$ and by 
keeping the constant leading term in the expansion of
the metric $g_{\mu\nu}(x)$ around the final point $x^{\mu}_f$,
thus making the space flat.
Obviously, one could also use the exact solution of the classical
equations of motion as background trajectory, but this would not change 
the result of the computation. It would correspond just to a different 
parametrization of the space of paths.

The quantum fields $q^\mu (\tau)$ in (\ref{para})
should vanish at the time boundaries 
since the boundary conditions are already included in $x^\mu_{bg}(\tau)$.
Therefore they can be expanded in a sine series. 
For the Lee-Yang ghosts we use the same Fourier 
expansion since the classical solutions of their field 
equations are $a^{\mu}=b^{\mu}=c^{\mu}=0$.
Hence
\be 
\phi^\mu(\tau) =\sum_{m=1}^{\infty} \phi^\mu_m \sin (\pi  m \tau)
\ee
where $\phi$ stands for all the quantum
fields $q^\mu, a^\mu, b^\mu, c^\mu $.
The measure $\tilde {\cal D}x $ in (\ref{meas}) is
now properly defined in  terms of integration over the Fourier 
coefficients $ \phi^\mu_m$ as follows
\be
\tilde {\cal D}x =
{\cal D}q {\cal D}a {\cal D}b {\cal D}c = 
\lim_{M \rightarrow \infty}
A \prod_{m=1}^{M}
\prod_{\mu=1}^D m \,
d q^\mu_m  d a^\mu_m  d b^\mu_m d c^\mu_m, \label{mes}
\ee
where $A$ is a constant.
Note that this fixes the path integral for a free particle to 
\be
\int \tilde {\cal D} x
\ \exp \biggl [ - { 1\over \beta} S_{free} \biggr ] = A \
{\rm e}^{-{\xi^2 \over {2\beta}}} \  
\label{mes2} 
\ee
where
\be 
S_{free} = \int_{-1}^{0}  \!\!\! d\tau\
{1\over 2} \delta_{\mu\nu} ( \dot x^\mu  \dot x^\nu
+ a^\mu a^\nu + b^\mu c^\nu) .
\ee
It is well-known that $A= (2\pi \beta)^{-{D\over2}}$,
however this value can also be deduced later on from a consistency
requirement. 

The way to implement mode regularization is now quite clear:
limiting the integration over the number of modes 
for each field to a finite mode number $M$ gives the natural
regularization of the path integral.
This regularization resolves the
ambiguities that show up in the continuum limit.

The perturbative expansion is generated by splitting the action into
a quadratic part $S_2$,  which defines the propagators, and an
interacting part $S_{int}$, which gives the vertices.
We do this splitting by expanding the action about the final point 
$x^\mu_f$ and obtain
\be
S = S_2 + S_{int} = S_2 + S_3 + S_4 + \dots
\ee
where
\bea
S_2 &=& \int_{-1}^{0}  \!\!\! d\tau\  {1\over 2}
g_{\mu\nu}(\xi^\mu \xi^\nu +
\dot q^\mu \dot q^\nu + a^\mu a^\nu +b^\mu c^\nu)
\\
S_3 &=& \int_{-1}^{0}\!\!\!  d\tau\ {1\over 2}
\partial_\alpha  g_{\mu\nu} (q^\alpha -\xi^\alpha \tau)
(\xi^\mu \xi^\nu + \dot q^\mu \dot q^\nu + a^\mu a^\nu +b^\mu c^\nu
- 2 \dot q^\mu  \xi^\nu)
\\
S_4 &=& \int_{-1}^{0}  \!\!\! d\tau \ \biggl [
{1\over 4} \partial_\alpha \partial_\beta g_{\mu\nu}
(q^\alpha q^\beta + \xi^\alpha \xi^\beta \tau^2 - 2 q^\alpha \xi^\beta \tau)
(\xi^\mu \xi^\nu + \dot q^\mu \dot q^\nu + a^\mu a^\nu 
+ b^\mu c^\nu -2 \dot q^\mu \xi^\nu) 
\nonumber \\
&& \hskip 1.3cm 
+\beta^2 (V+V_{MR})
\biggr ]. 
\eea
In this expansion all geometrical quantities, like $g_{\mu\nu}$ and
$\partial_\alpha  g_{\mu\nu}$, as well as $V$ and  $V_{MR}$,
are evaluated at the final point $x^\mu_f$,
but for notational simplicity we do not exhibit this dependence.
$S_2$ is taken as the free part and defines the propagators
which are easily obtained from the path integral
\bea
\la q^\mu(\tau) q^\nu(\sigma)\ra &=&
-\beta\ g^{\mu\nu}(x_f)\ \del(\tau,\sigma)
\nonumber\\
\la a^\mu(\tau) a^\nu(\si)\ra &=&  \beta\ g^{\mu\nu}(x_f)\
\dddel(\tau, \sigma) \label{propag}
\\
\la b^\mu(\tau) c^\nu(\sigma)\ra &=& -2\beta\ g^{\mu\nu}(x_f)\
\dddel(\tau,\sigma)
\nonumber
\eea
where
$\Delta$ is regulated by the mode cut-off 
\be
\Delta (\tau,\sigma) = \sum_{m=1}^{M}
\biggl [ - {2 \over {\pi^2 m^2}} \sin (\pi m \tau)
\sin (\pi m \sigma) \biggr ]
\ee
and has the following limiting value for $M \rightarrow \infty$
\be 
\D(\tau,\sigma)=\tau(\si+1)\te(\tau-\si)+\si(\tau+1)\te(\si-\tau)\, .
\ee
Note that we indicate
$\ddel (\tau,\sigma)={\partial \over {\partial  \tau}} \del(\tau,\sigma) $,
\ $\deld (\tau,\sigma)={\partial\over{\partial \sigma}} \del(\tau,\sigma)$
and so on. 
Details on the properties of these functions are given in appendix A.

Now, the quantum perturbative expansion reads:
\bea
&& \langle x^\mu_f,t_f | x^\mu_i,t_i \rangle  =
\int \tilde {\cal D} x
\ \exp \biggl [ - {1\over \beta} S \biggr ]
= A\ {\rm e}^{-{1\over 2\beta }g_{\mu\nu}\xi^\mu\xi^\nu}
\langle {\rm e}^{- {1\over \beta} S_{int}} \rangle \nonumber\\
&& = A\ {\rm e}^{-{1 
\over {2\beta}}g_{\mu\nu}\xi^\mu \xi^\nu}
\biggl ( \biggl \langle 1 -
{1\over \beta }S_3 - {1\over \beta } S_4 + {1\over  2\beta^2}
S_3^2 \biggr
\rangle + O(\beta^{3\over2}) \biggr ). \label{pex}
\eea
where the brackets $\langle \cdots \rangle$ denote the averaging with 
the free action $S_2$, and amount to use the propagators given in
(\ref{propag}) in 
the perturbative expansion. Note that in the last line of the 
above equation we have kept
only those terms contributing up to two loops, i.e. up to $O(\beta)$,
by taking into account that $\xi^\mu \sim
O(\beta^{1\over2})$, as follows from the exponential appearing in the
last line of~(\ref{pex}) after one averages over $\xi^\mu$.
Note also that
having extracted the coefficient $A$ together with 
 the exponential of the quadratic action $S_2$
evaluated on the background trajectory implies that 
the normalization of the left over path integral is such that 
$\langle 1 \rangle =1$. 

Using standard Wick contractions and going through a lengthy calculation 
one gets \cite{eight}
\bea
\biggl \langle - {1\over \beta} S_3 \biggr \rangle &=& 
- {1\over \beta} {1\over 4}
\p_\alpha g_{\mu\nu} \xi^\alpha \xi^\mu \xi^\nu, \label{s3}
\\
\biggl \langle - {1\over \beta} S_4 \biggr \rangle &=&
\p_\alpha \p_\beta g_{\mu\nu} \biggl [ {\beta\over 24}
(g^{\mu\nu} g^{\alpha\beta} -g^{\mu\alpha} g^{\nu\beta})
-{1\over 24}(g^{\mu\nu} \xi^\alpha \xi^\beta +g^{\alpha\beta}
\xi^\mu \xi^\nu - 2 g^{\mu\alpha} \xi^\nu \xi^\beta ) \nonumber \\
&& -{1\over\beta}{1\over 12} \xi^\mu \xi^\nu \xi^\alpha \xi^\beta \biggr]
+\beta ( V+V_{MR}), \label{s4}\\
\biggl \langle {1\over 2\beta^2} S_3^2 \biggr \rangle &=&
\p_\alpha g_{\mu\nu} \p_\beta g_{\lambda\rho} \biggl [
{\beta \over 96} (g^{\alpha\beta} g^{\mu\nu} g^{\lambda \rho}
- 4 g^{\alpha\rho} g^{\mu\nu} g^{\beta\lambda}
- 6 g^{\alpha\beta} g^{\mu\lambda} g^{\nu\rho}
\nonumber \\ &&
+ 4 g^{\alpha\rho} g^{\beta\mu} g^{\nu\lambda}
+ 4 g^{\alpha \mu} g^{\beta\lambda} g^{\nu\rho} )
\nonumber \\ &&
+{1\over 48} \biggl(g^{\mu\lambda} g^{\nu\rho} \xi^\alpha \xi^\beta
+ 2( g^{\alpha\beta} g^{\mu\lambda} - g^{\alpha\lambda} g^{\mu\beta})
\xi^\nu \xi^\rho
+ (2 g^{\alpha\lambda} g^{\beta\rho}-  g^{\alpha\beta} g^{\lambda\rho})
\xi^\mu \xi^\nu
\nonumber\\ &&
+ (2  g^{\mu\beta} g^{\lambda\rho} - 4 g^{\mu\lambda} g^{\beta\rho})
\xi^\alpha \xi^\nu \biggr )
\nonumber\\ &&
+{1\over \beta} {1\over 96} (g^{\alpha\beta} \xi^\mu \xi^\nu \xi^\lambda
\xi^\rho
- 4 g^{\alpha\lambda} \xi^\mu \xi^\nu \xi^\beta \xi^\rho
+ 4 g^{\mu\lambda} \xi^\alpha \xi^\nu \xi^\beta \xi^\rho )
\nonumber\\ &&
+{1\over \beta^2}{1\over 32} \xi^\alpha \xi^\mu \xi^\nu \xi^\beta
\xi^\lambda \xi^\rho \biggr ] . \label{s32}
\eea
This gives the transition amplitude in the two-loop approximation. 

To test its consistency 
one can use it to evolve in time an arbitrary wave function $\Psi(x,t)$
\be
\Psi(x_f,t_f) = \int \! d^D \!x_i \sqrt{g(x_i)}
\langle x^\mu_f,t_f | x^\mu_i,t_i \rangle  \Psi(x_i,t_i) 
\label{test}
\ee
and verify if $\Psi(x_f,t_f)$
solves the correct Schr{\"o}dinger equation\footnote{
The factor $\sqrt{g(x_i)}$ appearing in eq. (\ref{test})
is suggested by the expression of the path integral in eq. (\ref{pi})
which is formally a scalar for $V_{MR}=0$.
However general coordinate invariance is broken
by mode regularization, and recovered thanks to the effects of 
the counterterm $V_{MR}$. Therefore the measure appearing in 
(\ref{test}) should be considered as an ansatz 
which is verified, for example, by the calculations
presented in the next chapter.}.
Since the transition amplitude~(\ref{pex}) 
together with the results
(\ref{s3}), (\ref{s4}) and  (\ref{s32}) 
is given in terms of an expansion around
the final point $ (x_f,t_f)$, one Taylor expands 
the wave function $ \Psi(x_i,t_i) $ and the measure
$ \sqrt{g(x_i)} $ in  eq.~(\ref{test}) 
about that point,
performs the integration over $d^D x^{\mu}_i=d^D\xi^{\mu}$, 
and matches the various terms. The leading term fixes $A$
\be
\Psi = A (2 \pi \beta)^{D\over2} \Psi
\ \ \ \ \ \rightarrow \ \ \ \ \
A = (2 \pi \beta)^{-{D\over2}}, \label{zero}
\ee
and the terms of order $\beta$ give
\be
\beta \biggl [ - \p_t \Psi + {1\over 2} \nabla^2 \Psi 
- V \Psi \biggr ]   = 0 .
\label{obeta}
\ee
This last equation means that the wave function $\Psi$ 
satisfies the correct Schr{\"o}dinger equation (\ref{schr})
at the final point  $(x^\mu_f, t_f)$.

It is interesting to note that the counterterm $V_{MR}$
appears only in the last line of eq. (\ref{s4}).
Actually the value of the counterterm reported in eq. (\ref{ct})
has been deduced in \cite{eight}
by imposing that the transition amplitude would solve 
eq. (\ref{obeta}). General arguments can then be used to show that this
counterterm should be left unmodified at higher loops.
In fact one can consider quantum mechanics on curved spaces as
a super-renormalizable one-dimensional quantum field theory,
and check by power counting that all possible divergences can only appear 
at loop order 2 or less in $\beta$.
In the next section we are going to check that it is so indeed,
expelling doubts which have sometimes been raised
that mode regularization would be inconsistent at higher loops.
Thus one can consider mode regularization as a viable way of 
correctly defining the path integral in curved spaces.

\section{The transition amplitude at three loops}

In this section we want to check eq. (\ref{test})
at the next order in $\beta$, which is equivalent to showing that the 
transition amplitude  computed by the path integral
satisfies the Schr{\"o}dinger equation
not only at the point $(x^\mu_f,t_f)$ but in a small neighbourhood of it.
This computation can be quite lengthy if done in arbitrary coordinates.
To make it feasible we select a useful set of coordinates:
the Riemann normal coordinates centred at the point  $x^\mu_f$.
In such a frame of reference the coordinates
of an arbitrary point $x^\mu$ contained in a neighbourhood of the origin 
are given by a vector $z^\mu(x)$ belonging to the tangent
space at the origin. This vector
specifies the unique geodesic connecting 
the origin to the given point $x^\mu$ in a unit time.
In such a frame of reference the coordinates of the origin are obviously 
given by $z^\mu(x_f) = 0$.
In what follows we will use Riemann normal coordinates 
which we keep denoting by $x^\mu$ since no confusion can arise.

The expansion of the metric around the origin 
is given by (see e.g. \cite{seven} for a derivation)
\bea
 g_{\mu\nu}(x) &=&
g_{\mu\nu}(0) + {1\over 3} R_{\alpha\mu\nu\beta}(0) x^\alpha x^\beta
+{1\over 6} \nabla_\gamma R_{\alpha\mu\nu\beta}(0)
x^\alpha x^\beta x^\gamma 
\nonumber
\\
&& + \biggl(
{1\over{20}} \nabla_\gamma \nabla_\delta R_{\alpha\mu\nu\beta}(0) +
{2\over{45}} R_{\alpha\mu\sigma\beta} R_{\gamma\nu}{}^\sigma{}_\delta (0)
\biggr ) x^\alpha x^\beta x^\gamma x^\delta + O(x^5).
\eea
Note that the coefficients in this expansion are tensors
belonging to the tangent space at the origin.
This is a property of Riemann normal coordinates.

In general, the terms contributing to the transition amplitude up to 
three loops are given by
\bea
&& \langle x^\mu_f,t_f | x^\mu_i,t_i \rangle  
= A {\rm e}^{-{1\over 2\beta }g_{\mu\nu}\xi^\mu\xi^\nu}
\langle {\rm e}^{- {1\over \beta} S_{int}} \rangle 
\nonumber
\\
&& = A {\rm e}^{-{1\over {2\beta}} g_{\mu\nu}\xi^\mu \xi^\nu}
\biggl ( \biggl \langle 1 -
{1\over \beta }(S_3 + S_4 + S_5 + S_6) + 
{1\over  2\beta^2} ( (S_3 + S_4)^2 + 2 S_3 S_5 )
\nonumber \\
&& -{1\over  6\beta^3} ( S_3^3 + 3 S_3^2 S_4 )
 + {1\over  24\beta^4} S_3^4
\biggr \rangle + O(\beta^{5\over2}) \biggr ). \label{pex2} 
\eea
Clearly the computation would be quite complex in arbitrary coordinates.
Fortunately, 
in Riemann normal coordinates many terms are absent 
since we obtain
\bea
S_3 &=& 0 \\
S_4 &=& \int_{-1}^{0}   \!\!\! d\tau \ \biggl [
{1\over 6} R_{\alpha\mu\nu\beta} x^\alpha x^\beta
(\dot x^\mu \dot x^\nu + a^\mu a^\nu +b^\mu c^\nu) + 
\beta^2  ( V + V_{MR} ) \biggr ] 
\\
S_5 &=& \int_{-1}^{0}  \!\!\!  d\tau \ \biggl [
{1\over 12} \nabla_\gamma R_{\alpha\mu\nu\beta}
x^\alpha x^\beta x^\gamma 
(\dot x^\mu \dot x^\nu + a^\mu a^\nu +b^\mu c^\nu)
+ \beta^2  x^\alpha \partial_\alpha  ( V+ V_{MR} ) \biggr ] 
\\
S_6 &=& \int_{-1}^{0}   \!\!\!  d\tau \  \biggl [
\biggl(
{1\over{40}} \nabla_\gamma \nabla_\delta R_{\alpha\mu\nu\beta} +
 {1\over{45}} R_{\alpha\mu\sigma\beta} R_{\gamma\nu}{}^\sigma{}_\delta
\biggr )
x^\alpha x^\beta x^\gamma x^\delta 
(\dot x^\mu \dot x^\nu + a^\mu a^\nu +b^\mu c^\nu)
\nonumber
\\ 
&& \hskip 1.3cm 
+{\beta^2\over 2} x^\alpha x^\beta \partial_\alpha
 \partial_\beta  (V+ V_{MR} ) \biggr ] .
\eea
Note that all structures like $R_{\mu\nu\alpha\beta}$, $V$, $V_{MR}$
and derivatives thereof are evaluated at the origin of the Riemann
coordinate system, but for notational simplicity we do not indicate
so explicitly.  
The computation is still quite lengthy and we get
\bea
\biggl 
\langle - {1\over \beta} S_4 \biggr \rangle &=& 
{1\over 6} R_{\alpha\beta} \xi^\alpha \xi^\beta \ I_1 
- {\beta\over 6} R \ I_2 -\beta ( V + V_{MR}), 
\label{3s4}
\\
\biggl \langle - {1\over \beta} S_5 \biggr \rangle &=& 
- {1\over 12} \nabla_\gamma R_{\alpha\beta}
\xi^\alpha \xi^\beta \xi^\gamma  \ I_3 +
{\beta\over 6}\nabla_\alpha R \xi^\alpha \ I_4
-{\beta\over 2} 
\partial_\alpha ( V +V_{MR}) 
\xi^\alpha ,
\label{3s5}
\\
\biggl \langle - {1\over \beta} S_6 \biggr \rangle &=& \biggl 
({1\over 40}  \nabla_\gamma \nabla_\delta R_{\alpha\beta} +
{1\over 45} R_{\alpha\mu\nu \beta} R_\gamma{}^{\mu\nu}{}_\delta \biggr )
\xi^\alpha \xi^\beta \xi^\gamma \xi^\delta  \ I_5
-{\beta\over 40}\nabla^2 R_{\alpha\beta}
\xi^\alpha \xi^\beta \ I_6 \nonumber \\ 
&-&
\beta \biggl ({1\over 20}  (\nabla_\alpha \nabla_\beta R
+ \nabla^\mu \nabla_\alpha R_{\beta\mu} )
+{2\over 45} R_{\alpha\mu\nu \beta} R^{\mu\nu} \biggr)
\xi^\alpha \xi^\beta \ I_7 \nonumber \\
&-&
\beta \biggl ({1\over 20}  \nabla^\mu \nabla^\nu R_{\alpha\mu\nu\beta} +
{1\over 45} R_{\alpha\mu} R_\beta{}^\mu \biggr )
\xi^\alpha \xi^\beta \ (I_6 - I_7)
\nonumber \\
&-&
 {\beta\over 30} R_{\alpha\mu\nu \lambda} R_\beta{}^{\mu\nu \lambda} 
\xi^\alpha \xi^\beta \ (I_6 + I_7) 
+\beta^2 \biggl ({1\over 20}
\nabla^2 R + {1\over 45} R_{\mu\nu}^2
+ {1\over 30} R_{\alpha\mu\nu\beta}^2 \biggr )\ I_{8} 
\nonumber \\ 
&-& 
{\beta\over 6}\partial_\alpha \partial_\beta
(V + V_{MR}) \xi^\alpha \xi^\beta
+ 
{\beta^2\over 2}\partial^\alpha \partial_\alpha
(V + V_{MR})\ I_{9} , 
\label{3s6}
\\
\biggl \langle {1\over 2 \beta^2} S_4^2 \biggr \rangle &=&
{1\over 2} \biggl \langle - {1\over \beta} S_4 \biggr \rangle^2 +
\biggl \langle {1\over 2 \beta^2} S_4^2 \biggr \rangle_{\! con}, 
\label{3s41new}
\\
\biggl \langle {1\over 2 \beta^2} S_4^2 \biggr
\rangle_{\! con}
&=& 
 {1\over 72}  R_{\alpha\mu\nu\beta} R_\gamma{}^{\mu\nu}{}_\delta
\xi^\alpha \xi^\beta \xi^\delta \xi^\gamma \ I_{10}
-{\beta\over 72}  R_{\alpha\mu}  R_\beta{}^\mu \xi^\alpha \xi^\beta 
\ 4\ I_{11} \nonumber \\
&-&
{\beta\over 72} R_{\alpha\mu\nu\beta} R^{\mu\nu}
\xi^\alpha \xi^\beta \ 4 \ I_{12}
- {\beta \over 72} R_{\alpha\mu\nu\lambda}
R_{\beta}{}^{\mu\nu\lambda}\xi^\alpha \xi^\beta \ 
 6\ I_{13}
\nonumber \\
&+& {\beta^2 \over 72} R_{\mu\nu}^2\ 2 \ I_{14}
+ {\beta^2 \over 72}  R_{\alpha\mu\nu\beta}^2\ 3 \ I_{15} ,
\label{3s42}
\eea
where the integrals $I_n$ are listed and evaluated using
mode regularization in appendix A.
Inserting the specific values of the terms arising from 
the effective potential $V_{MR}$
when evaluated at the origin
\bea
&& V_{MR} = {1\over 8} R 
\\
&& \partial_\alpha V_{MR} = {1\over 8} \nabla_\alpha R 
\\
&& \partial_\alpha \partial_\beta V_{MR} = {1\over 8} 
\nabla_\alpha \nabla_\beta R -{1\over 36}
R_{\alpha\mu\nu\lambda}R_{\beta}{}^{\mu\nu\lambda}
\eea
leads us to the following expression for the transition amplitude at the 
third loop order
\bea
&& \langle x^\mu_f,t_f | x^\mu_i,t_i \rangle  
= {1\over (2 \pi \beta)^{{D\over 2}}}
{\rm e}^{-{1\over 2\beta }g_{\mu\nu}\xi^\mu\xi^\nu}
\biggl [ 
1 - {1\over 12} \xi^\alpha \xi^\beta R_{\alpha\beta}
-\beta \biggl( {1\over 12} R + V \biggr )
-{1\over 24} \xi^\alpha \xi^\beta \xi^\gamma \nabla_\gamma R_{\alpha\beta}
\nonumber 
\\
&&
-{1\over 2} \beta \xi^\alpha 
\nabla_\alpha  \biggl( {1\over 12} R + V  \biggr ) 
+
\xi^\alpha \xi^\beta \xi^\gamma \xi^\delta  
\biggl ( 
{1\over 360}   R_{\alpha\mu\nu \beta} R_\gamma{}^{\mu\nu}{}_\delta
+{1\over 288} R_{\alpha\beta} R_{\gamma\delta} 
-{1\over 80}  \nabla_\gamma \nabla_\delta R_{\alpha\beta}
\biggl ) 
\nonumber 
\\ 
&&
+\beta  \xi^\alpha \xi^\beta 
\biggl ( 
{1\over 360}  R_{\alpha\mu\nu\lambda} R_\beta{}^{\mu\nu\lambda} 
-{1\over 720} R_{\alpha\mu\nu\beta} R^{\mu\nu}
-{1\over 720} R_{\alpha\mu} R_\beta{}^\mu
+{1\over 12} \biggl( {1\over 12} R + V  \biggr ) R_{\alpha\beta} 
\nonumber 
\\
&& 
\hskip 1.5cm
-{1\over 240} \nabla^\mu \nabla^\nu R_{\alpha\mu\nu\beta}
-{7\over 480} \nabla_\alpha \nabla_\beta R 
-{1\over 6} \nabla_\alpha \nabla_\beta V 
\biggr )
\nonumber 
\\
&&
+ \beta^2 \biggl ( 
{1\over 720} R_{\alpha\mu\nu\beta}^2 
-{1\over 720} R_{\alpha\beta}^2
+{1\over 2} \biggl( {1\over 12} R + V  \biggr )^2
-{1\over 120} \nabla^2 R 
-{1\over 12} \nabla^2 V
\biggr ) 
+ O(\beta^{5\over 2}) \biggr ].
\label{resuno}
\eea
This is the complete expression which should be used to test
eq.~(\ref{test}) at order $\beta^2$.
A straightforward calculation shows that one indeed obtains 
an identity after making use of eq.~(\ref{obeta}).
The mode regulated path integral described
in the previous section passes this consistency check.
Therefore it can be considered as a well defined way of 
computing path integrals in curved spaces.

Before closing this section it may be useful to cast the transition
amplitude in a more compact form which can be made manifestly
symmetric under the exchange of the initial and final point.
Keeping on using the Riemann normal coordinates (in which we recall 
$x^\mu_f = 0 $ and $\xi^\mu \equiv x^\mu_i- x^\mu_f = x^\mu_i$)
and defining symmetrized quantities as
\be
\overline{ A} = {1\over 2} [A(x_i)  + A (x_f)]
\ee
we can write
\bea
&& \langle x^\mu_f,t_f | x^\mu_i,t_i \rangle  
= {1\over (2 \pi \beta)^{{D\over 2}}}
\exp \biggl [ 
-{1\over 2 \beta }\xi^\mu\xi^\nu \overline{ g_{\mu\nu}}
-{1\over 12} \xi^\alpha \xi^\beta \overline{R_{\alpha\beta}}
-\beta \biggl ({1\over 12} \overline{ R} + \overline{ V }\biggr )
\nonumber
\\
&&
+ \xi^\alpha \xi^\beta \xi^\gamma \xi^\delta 
\biggl (
{1\over 360} \overline{ R_{\alpha\mu\nu \beta} R_\gamma{}^{\mu\nu}{}_\delta} +
{1\over 120} \overline{\nabla_\gamma \nabla_\delta R_{\alpha\beta}}
\biggr ) 
\nonumber
\\
&&
+\beta \xi^\alpha \xi^\beta 
\biggl ( 
{1\over 360} \overline{ R_{\alpha\mu\nu\lambda} R_\beta{}^{\mu\nu\lambda}} 
-{1\over 720} \overline{  R_{\alpha\mu\nu\beta} R^{\mu\nu}}
-{1\over 720} \overline{  R_{\alpha\mu} R_\beta{}^\mu}
\nonumber
\\
&& \hskip 1.5cm
-{1\over 240} \overline{ \nabla^\mu \nabla^\nu R_{\alpha\mu\nu\beta}}
+{1\over 160} \overline{ \nabla_\alpha \nabla_\beta R} 
+{1\over 12} \overline{ \nabla_\alpha \nabla_\beta V} 
\biggr ) 
\nonumber
\\
&&
+ \beta^2 
\biggl ( 
{1\over 720} \overline{ R_{\alpha\mu\nu\beta}^2} 
- {1\over 720} \overline{ R_{\alpha\beta}^2}
- {1\over 120} \overline{ \nabla^2 R} -{1\over 12} \overline{ \nabla^2 V}
\biggr ) +  O(\beta^{5\over 2}) \biggr].
\label{resdue}
\eea
From this expression one can extract 
(by re-expanding part of the exponential)
the leading terms of the Seeley-De Witt
coefficients $a_0,a_1,a_2$ for non-coinciding points and
obtain, in particular,
the one loop trace anomalies for the operator 
$H= - {1\over 2 } \nabla^2  + V(x)$ in two and four dimensions.

\section{Time discretization}

The computation performed in the previous section was cast 
in such a way that can be easily extended to a different regularization 
scheme:
the time discretization method developed in refs. \cite{nine}.
Such a regularization was obtained by deriving directly from operatorial
methods a discretized version of the path integral.
Taking the continuum limit one recognizes the action with the proper 
counterterm, and the rules how to compute Feynman graphs.
These rules differ in general from the one required by mode 
regularization. The counterterm $V_W$ arising in
time discretization differs from $V_{MR}$, too.

The time discretization method leads to the following path integral
expression of the transition amplitude \cite{nine}
\be
\langle x^\mu_f,t_f | x^\mu_i,t_i \rangle  =
A \biggl [ {g(x_f)\over g(x_i)} \biggr ]^{1/4}
{\rm e}^{-{1\over 2\beta }g_{\mu\nu}(x_f)\xi^\mu\xi^\nu}
 \langle {\rm e}^{-{1\over \beta}S_{int}}
 \rangle,
\label{pitd}
\ee
where
\bea
&&\hskip -1cm
S_{int}=
\int_{-1}^{0} \!\!\! 
d\tau \ \biggl [
{1\over 2} \biggl ( g_{\mu\nu}(x) -g_{\mu\nu}(x_f) \biggr ) 
(\dot x^\mu \dot x^\nu + a^\mu a^\nu +
b^\mu c^\nu) + \beta^2 \biggl ( V(x) + V_{W}(x) \biggr ) \biggr ]
\label{quactd}  \\
&& \hskip -1cm
 V_{W} = 
{1\over 8} R  +{1\over 8} g^{\mu\nu} 
\Gamma_{\mu\alpha}{}^\beta \Gamma_{\nu\beta}{}^\alpha
\label{cttd} 
\\
&&\hskip -1cm
A= (2 \pi \beta)^{-{D\over2}} .
\eea
The propagators to be used in the perturbative expansion implied 
by the brackets on the right hand side of eq. (\ref{quactd}) are the 
same as in (\ref{propag}). The only difference is in the prescription 
how to resolve the ambiguities arising when distributions are multiplied
together. The prescription imposed by time discretization
consists in integrating the Dirac delta functions 
coming form the velocities and the ghosts 
propagators (thanks to the Lee-Yang ghosts they never appear 
multiplied together) and using consistently the value $\theta(0)={1\over2}$
for the step function. Note also the presence of the factor 
$ [ {g(x_f)\over g(x_i)} ]^{1/4}$ appearing in this scheme.

The result of the calculation has the same structure as 
the one reported in eqs.
(\ref{3s4}), (\ref{3s5}), (\ref{3s6}), (\ref{3s41new}),
(\ref{3s42}) 
with the difference that $V_{MR}$ should 
be substituted by $V_W$, leading to
\bea
&& V_{W} = {1\over 8} R 
\\
&& \partial_\alpha V_{W} = {1\over 8} \nabla_\alpha R 
\\
&& \partial_\alpha \partial_\beta V_{W} = {1\over 8} 
\nabla_\alpha \nabla_\beta R -{1\over 24}
R_{\alpha\mu\nu\lambda}R_{\beta}{}^{\mu\nu\lambda},
\eea
and with the following different values of the integrals computed 
in time discretization regularization
\be
I_1=0,\ \ \
I_3=0,\  \ \ 
I_5=0,\ \ \ 
I_{10}=0,\ \ \ 
I_{13}= {1\over 12},\ \ \ 
I_{15}= -{1\over 12}.
\ee
The other integrals are as in mode regularization.
Inserting all these values back in (\ref{pitd})
and expanding the coefficient 
$ [ {g(x_f)\over g(x_i)} ]^{1/4}$ 
at the required loop order give the same transition amplitude  
as in (\ref{resuno}) or, equivalently, in (\ref{resdue}).
Thus this result constitutes a successful test on the method 
developed in \cite{nine}.

\section{Conclusions}

In this paper we have discussed a proper definition of the configuration
space path integral for a particle moving in curved spaces.
By performing a three loop computation we have tested its consistency
and checked that one can equally well obtain the perturbative solution
of the  Schr{\"o}dinger equation by path integrals.
This fills a conceptual gap, showing that the perturbative 
description of a quantum particle moving in a curved space
obtained by De Witt by solving the Schr{\"o}dinger equation 
(i.e. using the canonical formulation of quantum mechanics
\cite{one}) can equally well be obtained in the path integral approach
introduced by Feynman fifty years ago.
This approach may also have practical applications in quantum 
field theoretical computations when  carried out in curved background 
using the world line formalism \cite{last}.

We have mainly described the mode regulated path integral.
Its definition was obtained in 
\cite{six}, \cite{seven} and \cite{eight}
by using a pragmatic approach to identify its key elements, 
and needed a strong check to test its foundations.
This we have provided in this paper. We find that the method of 
mode regularization is also quite appealing for aesthetic
reasons, since it is close to the spirit of path integrals 
that are meant to give a global picture of the quantum phenomena.

On the other hand we have also extended our computation to the
time discretization method of defining the path integrals \cite{nine}.
This method is in some sense closer to the local picture
given by the differential Schr{\"o}dinger equation, since one
imagines the particle propagating by small time steps.
It is nevertheless a consistent way of defining the path integral,
maybe superior at this stage, since one obtains its properties 
directly from canonical methods.
As we have seen also this scheme gives the correct result
for the transition amplitude.

An annoying property of the two regularization schemes we have been 
discussing is that they both do not respect general coordinate invariance
in target space, 
and require specific non-covariant counterterms to restore that symmetry.
It would be interesting to find a reliable covariant regularization scheme
or, at least, a scheme which while breaking covariance 
(e.g. in the decomposition of the action into free 
and interacting parts) does not necessitates non-covariant counterterms.
\vskip 1.5cm

{\Large \bf Acknowledgements }
\vskip.7cm

We wish to thank P. van Nieuwenhuizen for discussions and careful 
reading of the manuscript and J. Zinn-Justin for discussions.
\newpage

\appendix
\section{Appendix}

The function $ \del(\tau,\sigma)$ appearing in the propagators
is given in mode regularization by
\be
\del(\tau,\sigma) = \sum_{m=1}^{M} \biggl [
- {2\over {\pi^2 n^2}} {\rm \sin}(\pi m\tau)
{\rm \sin}(\pi m\sigma)\biggr ]
\label{delta}
\ee
and leads to the following limiting values as
 $M\rightarrow\infty$, at least in the bulk
(we recall that $ -1\leq \tau,\sigma\leq 0$),
\beq 
&&
\del(\tau,\sigma) = 
\tau (\sigma + 1 ) \theta (\tau-\sigma)
+ \sigma (\tau + 1) \theta (\sigma-\tau) 
\\
&&
\ddel (\tau,\sigma) 
= 
{\partial \over {\partial  \tau}} \del(\tau,\sigma) 
= \sigma + \theta(\tau -\sigma) 
\\
&&
\deld (\tau,\sigma) 
=
{\partial \over {\partial  \sigma}} \del(\tau,\sigma)
= \tau + \theta(\sigma - \tau)
\\
&&
\ddeld (\tau,\sigma) 
=
{\partial \over {\partial  \tau}} {\partial \over {\partial  \sigma}} 
\del(\tau,\sigma) 
=  1 - \delta(\tau -\sigma) 
\\
&&
\dddel (\tau,\sigma) 
=
{\partial^2 \over {\partial  \tau^2}} \del(\tau,\sigma)
= \delta(\sigma - \tau) .
\eeq
It is also useful to report the following limiting values
for coinciding points
\beq
&& \del(\tau,\tau) = \tau^2 +\tau \\
&& \deld(\tau,\tau) = \ddel(\tau,\tau) = \tau + {1\over 2} .
\eeq
Note that at the regulated level one can easily obtain the
following identities by inspection of
(\ref{delta}) and its derivatives
\beq
&& 
\ddeld(\tau,\tau)+ \dddel(\tau,\tau)
= \p_\tau (\ddel(\tau,\tau)) \\
&& \deld(\tau,\tau)= 0 \ \ {\rm at}\ \tau=-1,0  \\
&& \partial_\tau (\del(\tau,\tau)) = 2 \deld(\tau,\tau) \\
&& \dddel(\tau,\sigma) = \deldd(\tau,\sigma).  
\eeq

The limiting values given above should be used with care in the 
perturbative expansion of the path integral. Rather, one should 
resort to the proper regularized expressions whenever ambiguities
arise. In mode regularization we have adopted the following strategy:
one should partially integrate
as much as possible to reach expressions which are free of ambiguities,
and which can be computed directly in the continuum limit.
Following this procedure we have obtained the following results needed 
in the text
\bea
I_1 &=&  
\int_{-1}^{0}  d\tau \ \bigl (\tau^2 \ (\ddeld + \dddel) +
\del - 2 \tau\ \ddel \bigr )|_\tau = - {1\over 2} 
\\
I_2 &=& 
\int_{-1}^{0} d\tau \ \bigl (
\del \ (\ddeld + \dddel) - \ddel^2 \bigr )|_\tau
= - {1\over 4} 
\\
I_3 &=& 
\int_{-1}^{0}  d\tau \ 
\bigl ( \tau^3 \ (\ddeld + \dddel) + \tau \ 
\del -2 \tau^2\  \ddel \bigr )|_\tau 
= {1\over 2} 
\\
I_4 &=& 
\int_{-1}^{0}  d\tau \ 
\bigl ( \tau \ (\ddeld + \dddel) - \tau \  \ddel^2 \bigr )|_\tau 
= {1\over 8} 
\\
I_5 &=& 
\int_{-1}^{0}  d\tau \  
\bigl (\tau^4 \ (\ddeld + \dddel) + \tau^2 \ \del -2 \tau^3\  \ddel 
\bigr )|_\tau  
= - {1\over 2} 
\\
I_6 &=& 
\int_{-1}^{0}  d\tau \ 
\bigl ( \tau^2 \ \del\ (\ddeld + \dddel)+ \del^2 -2 \tau\  
\ddel\ \del \bigr )|_\tau  
= 0
\\
I_7 &=&
\int_{-1}^{0}  d\tau \ 
\bigl ( \tau^2 \ \del\ (\ddeld + \dddel) -\tau^2 \  \ddel^2 \bigr )|_\tau  
= - {1\over 12} 
\\
I_{8} &=&
\int_{-1}^{0}  d\tau \ 
\bigl ( \del^2\ (\ddeld + \dddel) - \ddel^2\ \del \bigr )|_\tau  
=  {1\over 24} 
\\
I_{9} &=&
\int_{-1}^{0}  d\tau \ 
\del|_\tau  
=  -{1\over 6} 
\\
I_{10} &=&
\idtds \biggl (
2 \tau^2 \ (\ddeld{}^2 - \dddel^2 )\ \sigma^2 
+ 4 \tau^2 \ \ddel^2
-8 \tau^2 \ \ddel\ \ddeld\ \sigma 
\nonumber 
\\ 
&+& 
2\  \del^2 
-8\ \del\ \deld\ \sigma 
+4 \tau \ \del\ \ddeld\  \sigma
+ 4\tau\ \ddel \ \deld\ \sigma
\biggr )
= 1
\\
I_{11} &=&
\idtds \biggl(
\tau \ (\ddeld + \dddel)|_\tau \ \del \ (\ddeld + \dddel)|_\sigma \ \sigma
+
\tau \ \deld|_\tau \ \ddeld \ \deld|_\sigma \ \sigma
\nonumber 
\\ 
&-&
2\tau \ (\ddeld + \dddel)|_\tau \ \deld\  \deld|_\sigma \ \sigma 
+ \del|_\tau \ \ddeld \ \del|_\sigma
+ \deld|_\tau \ \del \ \deld|_\sigma 
\nonumber 
\\ 
&-&
2 \del|_\tau \ \ddel\ \deld|_\sigma 
+
2 \tau \ (\ddeld + \dddel)|_\tau \ \deld\ \del|_\sigma
+
2 \tau \ \deld|_\tau \ \ddel\ \deld|_\sigma
\nonumber \\ 
&-& 2 \tau \ (\ddeld + \dddel)|_\tau \ \del \  \deld|_\sigma
-
2 \tau\ \deld|_\tau \ \ddeld\  \del|_\sigma  \sigma
\biggr ) 
= - {1\over 12}
\\
I_{12} &=&
\idtds
\bigg (
\tau^2 \ (\ddeld{}^2 - \dddel^2 ) \ \del|_\sigma
+
\deld{}^2 \ \del|_\sigma
-2 \tau \ \deld\ \ddeld\ \del|_\sigma
\nonumber  \\ 
&+&
\tau^2 \ \ddel^2 \ (\ddeld +\dddel)|_\sigma
+
\del^2 \ (\ddeld + \dddel)|_\sigma
- 
2 \tau \ \del\ \ddel\  (\ddeld + \dddel)|_\sigma
\nonumber
\\ 
&-& 
2 \tau^2 \ \ddel\ \ddeld\ \deld|_\sigma
- 
2 \del\ \deld \ \deld|_\sigma
+ 
2\tau\  \del \  \ddeld\ \deld|_\sigma
\nonumber
\\
&+&
2 \tau \ \ddel\ \deld\ \deld|_\sigma
\biggr ) 
={1\over 6}
\\
I_{13} &=&
\idtds \biggl (
\tau\ \del\ (\ddeld{}^2 - \dddel^2)\  \sigma 
-
\tau\ \ddel\ \deld\ \ddeld\ \sigma
+
\del^2\ \ddeld 
\nonumber \\ 
&-&
\del\ \ddel \ \deld 
+
2 \tau\ \deld\ \ddel^2 
-
2 \tau\ \del\ \ddel\ \ddeld
\biggr ) 
=  {1\over 18}
\\
I_{14} &=&
\idtds
\biggl (
\del|_\tau \ (\ddeld{}^2 - \dddel^2 )\  \del|_\sigma
-
4\ \del|_\tau \ \ddeld\ \ddel\ \deld|_\sigma
\nonumber \\ 
&+&
2\ \del|_\tau \ \ddel^2\  (\ddeld + \dddel)|_\sigma
+
2\ \deld|_\tau\ \del\ \ddeld\  \deld|_\sigma 
+
2\ \deld|_\tau\ \ddel\ \deld\ \deld|_\sigma
\nonumber \\ 
&-&
4\ \deld|_\tau \ \del\ \ddel\ (\ddeld + \dddel)|_\sigma
+
(\ddeld + \dddel)|_\tau \ \del^2 \ (\ddeld + \dddel)|_\sigma
\biggr )
= - {1\over 12}
\\
I_{15}
&=&
\idtds \biggl (
\del^2 \ (\ddeld{}^2 - \dddel^2 )
+
\ddel^2 \ \deld{}^2
-2\ \del\ \ddel\ \deld\ \ddeld
\biggr )
= - {1\over 18}
\eea

On the other hand, the time discretization method needs a 
different prescription in order to resolve the ambiguities.
It consists in integrating the Dirac delta functions whenever they appear
(the Lee-Yang ghosts guarantee that they never appear 
multiplied together) and using consistently the value $\theta(0)={1\over2}$
for the step function. We present now a list of the elementary integrals 
needed in the text and whose values differ in the two regularizations.
We have reported both values, the one related to time discretization being
included in square brackets.

\beq
&&
\idt
\tau^2 \ 
(\ddeld + \dddel)|_\tau = -{1 \over 6} 
\hskip 2cm \biggl[ {1\over 3} \biggr]
\\
&&
\idt 
\tau^3 \ 
(\ddeld + \dddel)|_\tau = {1 \over 4}
\hskip 2cm \biggl[- {1\over 4} \biggr]
\\
&&
\idt 
\tau^4 \
(\ddeld + \dddel)|_\tau = -{3 \over 10}
\hskip 2cm \biggl[ {1\over 5} \biggr]
\\
&&
\idtds
\tau^2 \
(\ddeld{}^2 - \dddel^2 )\ 
\sigma^2  = {19\over 90}
\hskip 2cm \biggl[ -{13\over 45} \biggr]
\\ 
&&
\idtds
\tau\ \ddel\ \deld\ \ddeld\ \sigma
= - {1\over 36}
\hskip 2cm \biggl[- {1\over 18} \biggr]
\\ 
&&
\idtds
\del\ \ddel\ \deld\ \ddeld
={1\over 180}
\hskip 2cm \biggl[ {7\over 360} \biggr].
\eeq

\vfill\eject


\end{document}